\documentclass[conference]{IEEEtran}
\ifCLASSINFOpdf
\else
\fi
 \usepackage[caption=false,font=footnotesize]{subfig}

\usepackage[numbers]{natbib}
\usepackage{amsmath}
\usepackage{lipsum}
\usepackage{graphicx}
\usepackage{amssymb}

\DeclareMathOperator{\SEG}    {SampEn_G}

\makeatletter
\renewcommand{\IEEEauthorrefmark}[1]{\textsuperscript{#1}}
\makeatother

\begin{document}
%
\title{Sample entropy for graph signals: \\ An approach to nonlinear analysis of graph signals}

\author{
\IEEEauthorblockN{
Mei-San Maggie Lei\IEEEauthorrefmark{1,*},
John Stewart Fabila Carrasco\IEEEauthorrefmark{2},
Javier Escudero\IEEEauthorrefmark{1}
}
\IEEEauthorblockA{\IEEEauthorrefmark{1}Institute for Imaging, Data and Communications, School of Engineering, University of Edinburgh, UK}
\IEEEauthorblockA{\IEEEauthorrefmark{2}School of Computer Science and Informatics, Cardiff University, UK}
\IEEEauthorblockA{\IEEEauthorrefmark{*}Contacting Author, E-Mail: M.S.Lei@sms.ed.ac.uk.}
}



%


\maketitle

\begin{abstract}

We introduce a graph-signal generalisation of Sample Entropy, denoted SampEn$_{G}$, to quantify irregularity of graph signals on a continuous state space, complementing existing methods on symbolic dynamics. Our approach replaces the temporal delay embedding of classical SampEn with a multi-hop graph-based embedding: for each node, we aggregate patterns from local walk-weighted neighbourhood averages computed via powers of the graph shift operator. We show empirically that SampEn$_{G}$ reduces to classical 1D SampEn on directed path graphs, and validate its nonlinear sensitivity using the logistic map. Experiments on directed Erd\H{o}s--R\'enyi graph signals further characterise its behaviour with connectivity and pattern length $m$, with practical runtimes on the order of thousands of nodes. We expect SampEn$_{G}$ to open up new ways to analyse graph signals, generalising SampEn and the concept of conditional entropy to extending nonlinear analysis to a wide variety of network data.

\end{abstract}


%
\IEEEpeerreviewmaketitle


\section{Background}
Current entropy methods for Graph Signal Processing (GSP), including graph-based Permutation Entropy and its variants--Dispersion Entropy, and Bubble Entropy--have been extended to signals defined over graphs and networks \cite{fabila-carrascoPEG2022,FABILACARRASCO2023113977,bubbleen_graph}. While effective, they are restricted to symbolic dynamics through discretisation to compute Shannon entropy. 

Sample Entropy (SampEn) measures irregularity (complexity) of a time series by estimating a conditional recurrence probability in a continuous state space via a fixed threshold \(\epsilon\). Specifically, SampEn measures how likely patterns of length $m$ remains similar when extended to length $m+1$. Application of SampEn includes finance, electronics, ecology, and engineering \cite{ApproximateEntropySample}. 


In this abstract, we summarise our recent work extending SampEn to graph signal (\(\SEG\)), providing a continuous, conditional entropy-based entropy on graph signal framework. Our experiments with logistic map and Erd\H{o}s--R\'enyi (ER) graph signals then evaluate its sensitivity to nonlinear dynamics, and to study its dependence on connectivity using random node attributes.

\section{Methodology}
\subsection{SampEn for graph signals: SampEn$_{G}$}

Classical SampEn for univariate time series is controlled by the pattern length $m$ and tolerance parameter $r$, which sets the matching threshold \(\epsilon=r\times SD\) (standard deviation of signal), and the patterns are built by concatenating successive temporal observations separated by a time step ($L$ often taken as $L=1$) \cite{sampenfirstdefin}. In this abstract, we use the Chebyshev distance, with $m=2$ and $r=0.2$ , as is standard in SampEn \cite{sampen_typical_param}.

In our graph-based generalisation, the notion of a step is replaced by the graph hop --  one hop corresponds to a single traversal along an edge from one node to another adjacent node.

Consider a graph \(\mathcal{G} = (\mathcal{N}, \mathcal{E}, \mathbf{A})\). For each node, we construct a vector (pattern) whose components are local graph-signal aggregates to characterise the radial profiles around each node in multiples of $L$ hops ($L=0, 1, 2, \dots$) via powers of the graph shift operator (adjacency matrix) in analogy to increasing temporal steps in SampEn.

To obtain these components, we first identify the local $L$-hop neighbours for each node. Consider the multiplication of $\mathbf{A}$ with itself: 
\begin{equation}
    (\mathbf{A}^{L=2})_{ij} = \sum_{k} \mathbf{A}_{ik} \times \mathbf{A}_{kj}.
\end{equation}

Each entry $(\mathbf{A}^{L=2})_{ij}$ is the sum of walks of length \textit{two} from node $i$ to $j$ over any intermediate node $k$. The $L$-hop degree of node $i$ can be defined as the corresponding row sum:
\begin{align}
    \deg^{L}(i) &:= \sum_{j=1}^N (\mathbf{A}^L)_{ij}, \qquad L\ge 1.
    \label{eq:degL}
\end{align}

We identify the $L$-neighbourhood -- all nodes that are exactly at $L$-hops -- for each node. For $L\in\mathbb{N}$, non-zero entries in \((\mathbf{A}^L)_{ij}\) indicate the total number of $L$-length walks reachable from node $i$ to $j$. 

Building on our previous work on PE$_G$ \cite{fabila-carrascoPEG2022}, we here extend the framework to introduce SampEn$_{G}$. For each node \(i\in\mathcal{N}\), we construct its $m$ and $m+1$-dimensional patterns $\mathbf{\overline{x}}^{m}(i)$ and $\mathbf{\overline{x}}^{m+1}(i)$ as:
\begin{equation}
    \mathbf{\overline{x}}^{m}(i) = [\overline{x}_i^0, \overline{x}_i^1, \dots, \overline{x}_i^{m-1}] \quad i \in \mathcal{N}^\star_{m},
\end{equation}
\begin{equation}
    \mathbf{\overline{x}}^{m+1}(i) = [\overline{x}_i^0, \overline{x}_i^1, \dots, \overline{x}_i^{m}] \quad i \in \mathcal{N}^\star_{m},
\end{equation}
\begin{align}
    \mathcal{N}^\star_m := \big\{ i\in\mathcal{N} : \deg^{L}(i)>0 \quad \text{for}\, L=1,\dots,m \big\}.
    \label{eq:Vstar-final}
\end{align}

To avoid undefined patterns with sparse graphs, we restrict to patterns with non-zero reachability up to $m$-hops. Here, $\overline{x}_i^0$ is the signal value on node $i$, and $\overline{x}_i^L$ is a walk-weighted mean of the $L$-neighbourhood of node $i$, in hop-radius propagation:
\begin{equation}
    \quad \overline{x}_i^{L} = \frac{1}{\text{deg}^{L}(i)} \sum_{j \in \mathcal{N}_{L}(i)} (\mathbf{A}^{L})_{ij}x_j.
    \label{eq:segraphavg}
\end{equation}

\(\mathcal{N}_L(i)\) denotes the set of reachable nodes in \(\mathcal{N}\) by at least one walk of exactly length $L$ from node \(i\). For weighted graphs, we replace \(\mathbf{A}\) with the weighted adjacency \(\mathbf{W}\). The matrix power $\mathbf{W}^L$ is the total weighted walks of $L$ hops from $i$ to $j$: \(\text{deg}^{L}(i) = \sum_{j=1}^N (\mathbf{W}^L)_{ij}, 
    \quad \overline{x}_i^{L} = \frac{1}{\text{deg}^{L}(i)} \sum_{j \in \mathcal{N}_{L}(i)} (\mathbf{W}^{L})_{ij}x_j.
\)

The resulting patterns are of size $1\times m$ for each node. The successive steps of similarity matching, then follows:
\begin{equation}
    B_i(r) = \frac{1}{|\mathcal{N}^\star_{m}|-1}\!
    \sum_{\substack{j\in\mathcal{N}^\star_{m} \\ j\neq i}}
    \| d[\mathbf{\overline{x}}^{m}(i), \text{ } \mathbf{\overline{x}}^{m}(j)] \leq \epsilon \|,
    \label{eq:Bi-graph}
\end{equation}
where $\|\cdot\|$ denotes cardinality of a set, counting pairwise matches of patterns within Chebyshev distance of \(\epsilon=r\times SD\), the total sums are then computed to obtain $B^m(r)$: 
\begin{equation}
    B^m(r) = \frac{1}{|\mathcal{N}^\star_{m}|} \sum_{i\in\mathcal{N}^\star_{m}} B_i(r). 
    \label{eq:B_match_total_graph}
\end{equation}

Repeating for \(\mathbf{\overline{x}}^{m+1}(i)\) yields count of matches $A^m(r)$ for $m+1$-patterns. Finally, we compute SampEn as the ratio of $A^m$ and $B^m$:
\begin{equation}\small
    \text{SampEn}(m, r, N) = -\ln\left(\frac{A^m(r)}{B^m(r)}\right).
    \label{eq:sampenapproxfinal}
\end{equation}



\section{Experiments and Results}

\subsection{Logistic Map}\label{Sec:logistic}

We validated SampEn$_{G}$ on the logistic map, a classical nonlinear dynamical model that changes between order and chaos as a function of its bifurcation parameter $\rho$. The map is defined by \cite{logistic_map_sampen}: \begin{equation}
    x_{t+1} = \rho x_t (1 - x_t).
\end{equation}

Each node is associated with a time sample with directed \(i\rightarrow i+1\), or undirected edges between consecutive nodes, following \cite{fabila-carrascoPEG2022}. We computed SampEn$_{G}$ for both paths and evaluated them against classical 1D SampEn (implementation from \cite{MartinezCagigal2018SampleEntropy}) in relation to \(\rho\), repeated over 20 random initial conditions drawn uniformly from the open interval \((0,1)\). We report the mean and standard deviation across runs.

\begin{figure}[t]
    \centering
    \includegraphics[width=0.43\textwidth]{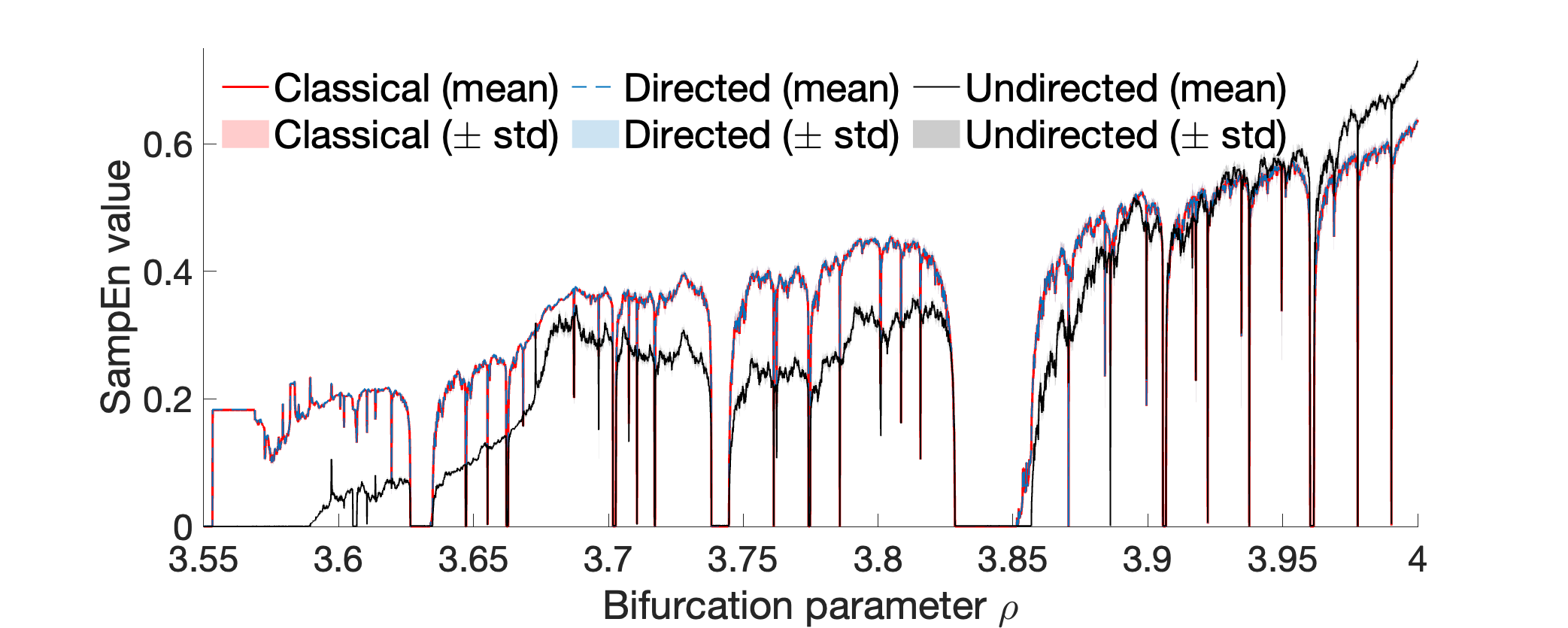}
    \caption{Directed (blue), undirected (black) SampEn\(_{G}\), classical SampEn (red) (\(m=2\), \(r=0.2\)) against $\rho$ for logistic map.}
    \label{fig:logisticmap}
\end{figure}

Fig.~\ref{fig:logisticmap} shows that SampEn$_{G}$ replicates and reduces to the standard 1D case on the directed path graphs. This is because the directed graph imposes the causality in the order of samples within the patterns of length $m$ and $m+1$ implicitly assumed by classical SampEn. The undirected-path SampEn$_{G}$ closely tracks the overall trend of the other two curves, correctly identifying the bifurcation points and islands of periodicity with slight quantitative differences due to the non-causal, symmetric neighbourhood structure of the undirected path. These results validate the generalisation property of SampEn$_{G}$ on 1D signals.

\subsection{Erd\H{o}s--R\'enyi graphs}

We considered binary, directed ER graphs of \(N=2700\). Each node value is a random sample drawn from a uniform distribution in the range $[0.01,0.10]$. The parameter $p$ controls the probability in which an edge exists between any two nodes. 

\begin{figure}[t]
        \centering
        \includegraphics[width=0.3\textwidth]{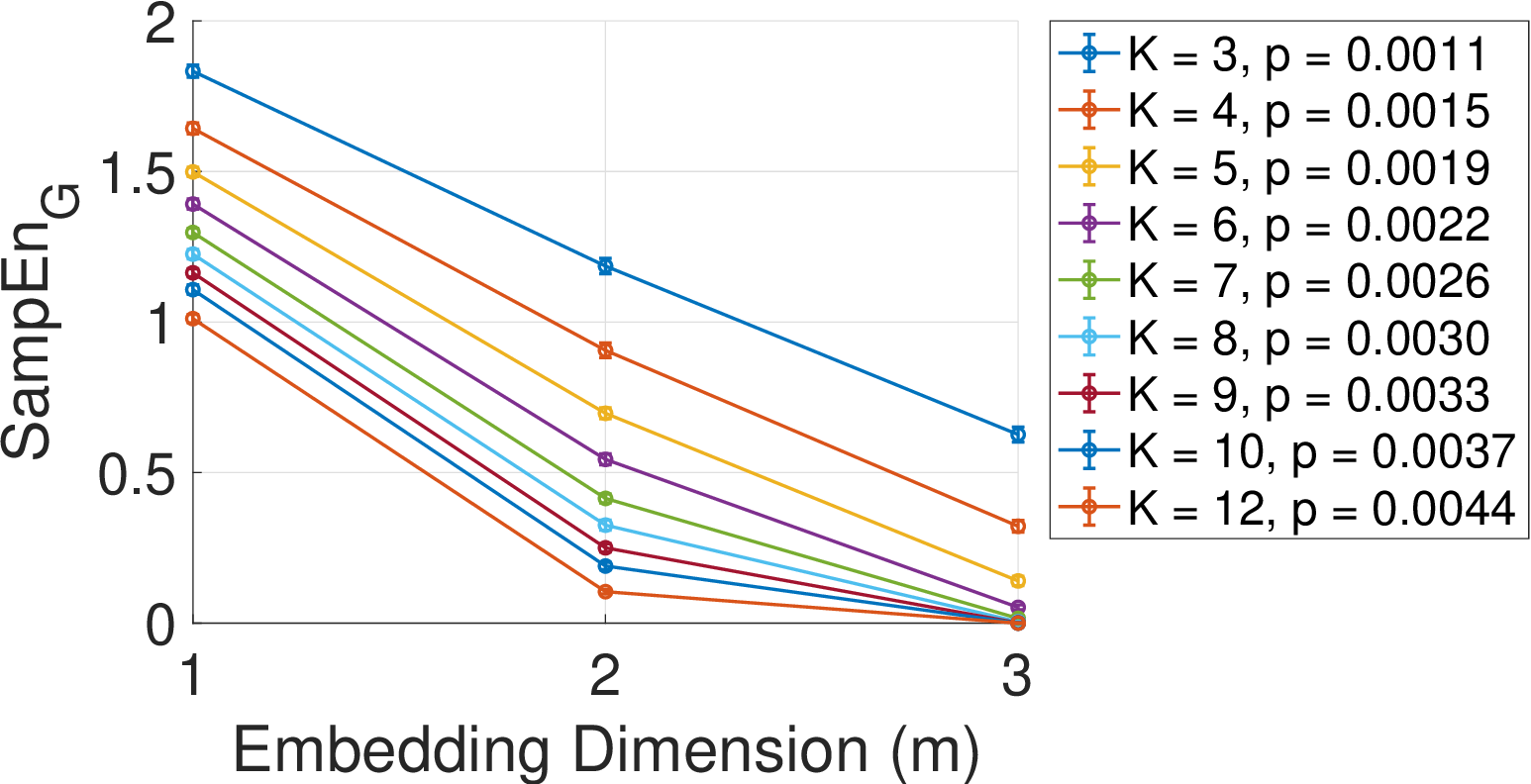}

    \caption{SampEn$_{G}$ on directed ER graphs ($ N=2700 $, $ r=0.2 $): mean$\pm$std vs target out-degree $ K $, connectivity $p$ for pattern length $m\in\{1,2,3\} $ over 20 realisations.}
    \label{fig:sampen_ER_res}

\end{figure}

We set \(
p = \frac{K}{N-1},
\) to target mean out-degree \(K\in\{3,4,5,6,7,8,9,10,12\}\) and varied pattern length $m \in [1,2,3]$. We then computed SampEn$_{G}$ over 20 graph realisations on an Apple M3 CPU running MATLAB R2024b. For \(N=2700\), we observed runtimes of $\sim295~ms$ ($m=1$) to $\sim1400~ms$ ($m=3$) per realisation. The resulting mean SampEn$_{G}$ across 20 observations are reported in Fig.~\ref{fig:sampen_ER_res}.

SampEn$_{G}$ is computationally practical. In our experiments, graphs of $N=2700$ nodes can be processed at approximately $1.4~s$ per run.

With sparse networks (small \(K\), low $p$), SampEn$_{G}$ decreases as connectivity $p$/$K$ increases, reflecting increased regularity in the topology. As the graphs become denser and introduce more long-range connections, SampEn$_{G}$ becomes less separable, and eventually approaches zero. This is more prominent with increased $m$, which substantially expands the outreach, and thus overlap, of local neighbourhoods, leading to homogeneity in patterns and reduces the effective variability across local patterns which SampEn$_{G}$ can exploit.



\section{Conclusion}

In this abstract, we introduced SampEn$_G$ as a topology-aware generalisation of Sample Entropy for graph signals using multi-hop graph embeddings. Experiments on the logistic map verified reduction to classical 1D SampEn with directed paths, and results confirmed sensitivity across the order–chaos transitions. On directed Erdős–Rényi graphs, SampEn$_G$ decreased with increasing connectivity, consistent with increased predictability and pattern recurrence. With denser graphs and larger $m$, SampEn$_G$ approached zero due to neighbourhood homogenisation induced by growing overlaps. Future work may explore alternative graph embeddings, robustness to noisy signal defined over graphs, and applications to real networked sensing domains.






%

\bibliographystyle{IEEEtranN}   
\bibliography{references_withouturl}






\end{document}